\documentclass[review,3p]{elsarticle}

\usepackage{amsmath,amssymb,amsfonts}
\usepackage{algorithmic}
\usepackage{graphicx}
\usepackage{textcomp}
\usepackage{soul}
\usepackage{float}
\usepackage{booktabs}
\usepackage{multirow}
\usepackage{subcaption}
\usepackage{array}
\usepackage{wrapfig}
\usepackage[utf8]{inputenc}
\usepackage{color, colortbl}
\usepackage{comment}
\usepackage{balance}
\usepackage[inline]{enumitem}
\usepackage{url}
\usepackage{adjustbox}

\definecolor{lightyellow}{cmyk}{0,0,0.5,0}
\sethlcolor{white}

\journal{Journal of Pervasive and Mobile Computing}

\begin{document}

\begin{frontmatter}

\title{Towards better social crisis data with HERMES:\\ Hybrid sensing for EmeRgency ManagEment System}

\author[unipi]{Marco Avvenuti}
\ead{marco.avvenuti@unipi.it}

\author[iit]{Salvatore Bellomo}
\ead{s.bellomo@iit.cnr.it}

\author[iit]{Stefano Cresci}
\ead{s.cresci@iit.cnr.it}

\author[unipi,iit]{Leonardo Nizzoli\corref{mycorrespondingauthor}}
\cortext[mycorrespondingauthor]{Corresponding author}
\ead{l.nizzoli@iit.cnr.it}

\author[iit]{Maurizio Tesconi}
\ead{m.tesconi@iit.cnr.it}

\address[unipi]{Dept. of Information Engineering, University of Pisa, Italy.}
\address[iit]{Institute of Informatics and Telematics, National Research Council (IIT-CNR), Italy.}

\begin{abstract}

People involved in mass emergencies increasingly publish information-rich contents in online social networks (OSNs), thus acting as a distributed and resilient network of human sensors. In this work we present HERMES, a system designed to enrich the information spontaneously disclosed by OSN users in the aftermath of disasters. HERMES leverages a mixed data collection strategy, called hybrid sensing, and state-of-the-art AI techniques. Evaluated in real-world emergencies, HERMES proved to increase:
\begin{enumerate*}[label=(\roman*)]
    \item the amount of the available damage information;
    \item the density (up to $7\times$) and the variety (up to $18\times$) of the retrieved geographic information;
    \item the geographic coverage (up to $30\%$) and granularity.
\end{enumerate*}
\end{abstract}


\end{frontmatter}

\section{Introduction}
\label{sec:intro}

The recent proliferation of mobile devices, equipped with a large array of sensors and communication capabilities, has established the so called Cyber–Physical convergence. In this high-tech scenario, information flows in a loop between the physical and the cyber worlds, mediated by human activities~\cite{conti2017internet, conti2012looking, ray2017internet}. At the same time, the mass diffusion and availability of online social networks (OSNs) has appointed them as the preferred supplier of information and communication services, especially during fast-paced, unfolding events that impose stringent time requirements~\cite{gao2014quantifying}. Crises and disasters are among such events and, indeed, many people involved in disasters publish information-rich textual and multimedia messages in OSNs, such as Facebook and Twitter, often live and \textit{in situ}~\cite{middleton2014real,avvenuti2017nowcasting,avvenuti2014ears, avvenuti2015pulling, solmaz2017modeling}.

The conjunction between the Cyber–Physical convergence and the rise of OSNs significantly extends, complements, and possibly substitutes, conventional sensing by enabling the collection of data through networks of humans. These unprecedented sharing and sensing opportunities have enabled situations where individuals not only play the role of sensor operators, but also act as data sources themselves, thus implementing the so-called Human-as-a-Sensor paradigm. 
This spontaneous behavior has driven a new thriving -- yet challenging -- research field, called social sensing, investigating how human-sourced data can be gathered and used to gain situational awareness in a number of socially relevant domains~\cite{cresci2018harnessing}.
Depending on their awareness and their involvement in the system, ``human sensors'' are faced with either \textit{opportunistic sensing}, where users spontaneously collect and share data that is transparently intercepted by a situation-aware system -- or with \textit{participatory sensing}, where users consciously meet an application request out of their own will~\cite{avvenuti2016framework}.

Challengingly, relevant information in OSNs is typically unstructured, heterogeneous and fragmented over a large number of messages, in such a way that it cannot be directly used. Hence, a number of AI techniques have been developed in order to turn that messy data into a small set of actionable, clear and concise messages~\cite{alam2017data, imran2015processing}. 
In the aftermath of mass-disasters, AI techniques are adopted to automatically process large-scale crisis data, in an effort to help tracking stricken locations, assessing the damage, coordinating the rescue efforts, and ultimately contributing to make communities stronger and more resilient~\cite{middleton2014real, avvenuti2016impromptu, avvenuti2016predictability, lin2017special, shah2019rising}.
Until now, these techniques have almost solely fed on those messages spontaneously shared in OSNs -- that is, by exclusively following an opportunistic sensing approach~\cite{cresci2019enriching}. 
In other words, in the last years we have mainly looked for ways for improving the AI techniques that make sense of social crisis data. However in sensing and in AI, better results are not only achieved via better algorithms, but also and foremost via more and better data~\cite{halevy2009unreasonable,domingos2012few}. Yet to date, almost no effort has been made towards developing sensing and computational methods for enriching available data in the first place -- e.g., by soliciting additional information from people directly involved in disasters.

\subsection{Contributions}
Following an orthogonal approach to previous endeavours in the fields of human sensing and AI for disaster management, we propose HERMES, a system 
for enriching available OSN data in the aftermath of disasters. HERMES automatically complements data spontaneously published and collected via opportunistic sensing, with targeted solicited data collected via participatory sensing.  
At a glance, the system listens to an OSN stream of messages related to a disaster. Then, it leverages AI techniques to select a subset of relevant OSN users directly involved in the disaster, from which to solicit additional information. Finally, it automatically asks targeted questions to selected users thus allowing to meet the information needs of Emergency Operation Centers (EOCs), and it collects answers to those questions in real-time.
By analyzing data collected by HERMES, we empirically demonstrate that such messages contain richer information with respect to those shared spontaneously (i.e., those typically used in previous works).
Detailed contributions are summarized in the following:
\begin{itemize}
 \item We design a novel sensing system, called HERMES, that implements the Human-as-a-Sensor paradigm and that exploits state-of-the-art AI techniques. By leveraging a hybrid sensing strategy, HERMES allows to combine the strengths of both opportunistic and participatory sensing.
 \item We experiment with HERMES in a practically relevant scenario -- such as emergency management -- demonstrating its usefulness towards the acquisition of more and better crisis data from human sensors.
 \item We report results of an extensive real-world experimentation during which we monitored 436 worldwide earthquakes. Our results show that HERMES allowed to increase the amount of the retrieved damage information; the density (up to $7\times$) and the variety (up to $18\times$) of the retrieved geographic information, and the geographic coverage (up to $30\%$) and granularity;
 \item \hl{We compare results obtained by HERMES with those of the well-established USGS service ``Did you feel it?'' (DYFI). We demonstrate HERMES capability in engaging a remarkably larger amount of contributors, with respect to USGS-DYFI, when considering earthquakes occurring outside of the U.S. and having little media coverage.}
\end{itemize}

Notably, HERMES leverages the Cyber-Physical information flow at its whole extent, embodying the paradigm of Cyber-Physical systems with humans-in-the-loop.

\subsection{Roadmap}
The remainder of this paper is organized as follows. The upcoming Section~\ref{sec:relwork} surveys relevant works in the fields of human sensing and AI for disaster management. In Section~\ref{sec:system} we design our proposed HERMES system; we describe its AI-powered logical components, \hl{and we reason about its possible legal and ethical implications}. Then, in Section~\ref{sec:results} we report and discuss results from the real-world experimentation campaign, \hl{and we compare our results with those of USGS-DYFI.} Finally, Section~\ref{sec:conc} draws conclusions and describes favorable avenues for future research and experimentation.
 \section{Related work in human sensing and AI for disaster management}
\label{sec:relwork}
Among the tasks typically performed to make sense of social crisis data, are: message filtering message classification according to the information conveyed, ranking of most relevant messages, aggregation and summarization of information, and extraction of relevant information mentioned in messages (i.e., people, places and objects involved in the disaster)~\cite{imran2015processing}.
These low-level tasks are usually carried out as part of more complex analyses, performed to achieve high-level goals such as increasing the overall situational awareness during a crisis, or maintaining an updated crisis map of the stricken area~\cite{yin2012using,middleton2014real}. Given the focus of our work, in this section we briefly survey previous solutions to the tasks of message filtering, message classification and information extraction, with the specific goal of producing crisis maps. A broader survey of recent literature can be accessed in~\cite{imran2015processing} and references therein.

Regarding message filtering and classification, the majority of existing approaches rely on natural language processing techniques~\cite{imran2013extracting,cresci2015linguistically,avvenuti2018realworld}. The task of message filtering is typically tackled with a binary classifier that labels messages as either relevant or irrelevant. Instead, message classification can take many different forms, according to the type of collected data and the goals of the analysis. For instance, authors of~\cite{cresci2015linguistically} classified messages as either conveying useful information for a damage assessment task or not. Other authors learned binary, multi-class, and multi-label classifiers for a wider set of goals, such as for identifying messages posted by witness and non-witness users~\cite{imran2013extracting,avvenuti2018realworld}, and messages conveying different types of information (e.g., cautions and advice, casualties and damage, donations)~\cite{imran2013extracting}. Conversely, a different approach to message classification is employed in the AIDR system~\cite{imran2014aidr}, where both human and machine intelligence are simultaneously employed for carrying out high-quality classification tasks in real-time.

Concerning information extraction, among the most useful information in the aftermath of a disaster are geographic references. The crucial importance of geospatial information, combined with the negligible number of messages that natively carry such information~\cite{avvenuti2018gsp}, resulted in many attempts to perform automatic geoparsing and geotagging of OSN data~\cite{zheng2018survey, imran2015processing}. Traditional approaches to geoparsing relied on pattern matching of message tokens against gazetteer data, or on named entity extraction and classification~\cite{middleton2014real}. Instead, more recent approaches adopted powerful AI algorithms for classification and semantic annotation, probabilistic language models and representations of structured information contained in knowledge graphs~\cite{avvenuti2018gsp}.

Finally, relevant information obtained via message classification or information extraction, is often leveraged to produce and update crisis maps, allowing the visualization of most relevant information about an event on a map~\cite{imran2015processing}. State-of-the-art systems produce crisis maps by comparing the volume of messages that mention given locations with statistical baselines~\cite{middleton2014real}, or by highlighting  regions of space for which the system collected a significantly high number of damage-related information~\cite{avvenuti2018crismap, cresci2015crisis}.

All works surveyed in this section are solely based on spontaneous user messages, opportunistically collected. As such, all these works could potentially benefit from a data collection strategy such as the one that we are proposing.

 \section{HERMES: Hybrid sensing for EmeRgency ManagEment System} 
\label{sec:system}

\begin{figure*}[ht!]
    \centering
    \includegraphics[width=\textwidth]{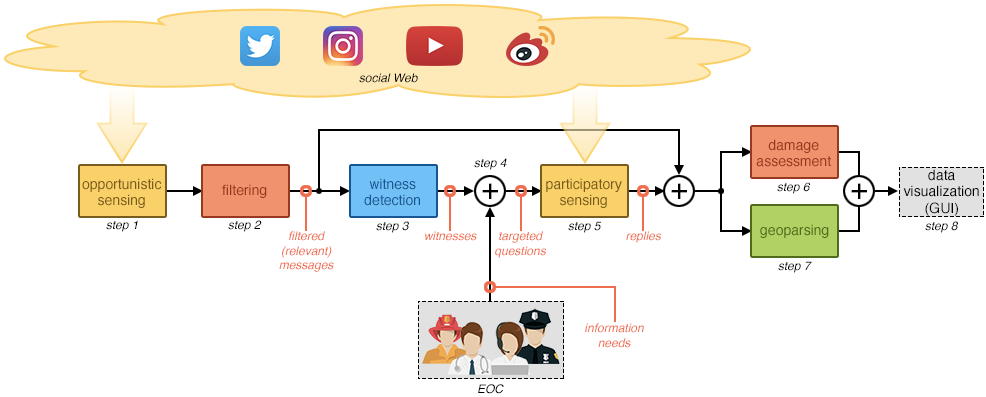}
    \caption{System architecture of HERMES, a Hybrid sensing for EmeRgency ManagEment System.}
\label{fig:system}
\end{figure*}

\begin{figure*}[ht!]
    \centering
    \includegraphics[width=0.5\textwidth]{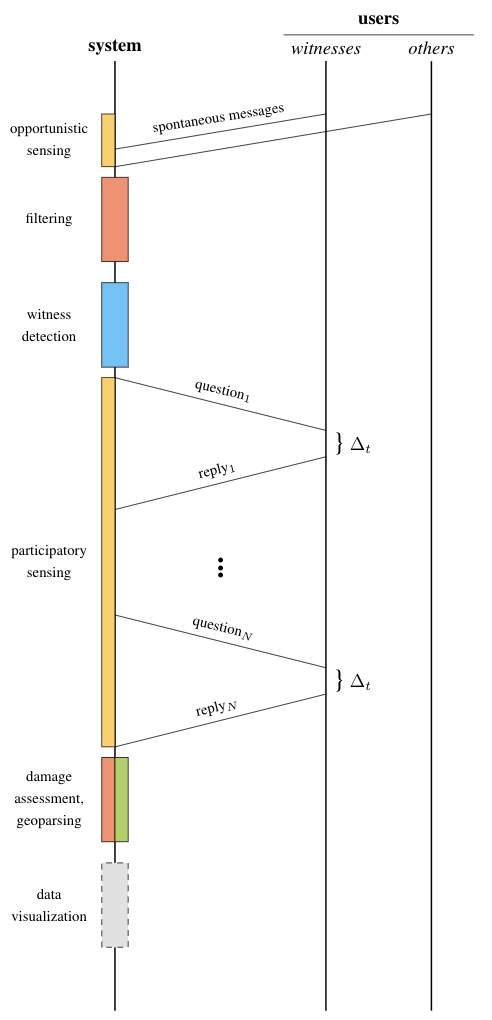}
    \caption{UML sequence diagram of the HERMES system. $\Delta_t$ represents the latency between the sending of our targeted questions and the possible user replies.}
    \label{fig:system-seqdiag}
\end{figure*}

\hl{In this section, we describe the high-level architecture of HERMES, a novel and fully-automatic emergency management system. HERMES combines state-of-the-art AI techniques for message filtering, damage assessment, geoparsing  and witness detection with a hybrid approach to sensing, taking advantage of both spontaneous and solicited information disclosed by OSN users.} \hl{A schema of the proposed system architecture is depicted in Figure~\mbox{\ref{fig:system}}, whereas Figure~\mbox{\ref{fig:system-seqdiag}} shows a temporal diagram of its functioning according to the Unified Modeling Language (UML) visualization approach. In the current implementation, HERMES focuses on earthquake emergency relief. Hence, its functioning is triggered by notifications automatically issued by the United States Geological Survey (USGS) whenever an earthquake occurs. However, we emphasize that our proposed approach can be easily generalized to other types of both natural and man-made crises such as floods, landslides as well as riots and terrorist attacks.} 

\hl{Once triggered, HERMES performs the following steps:}
\begin{enumerate}
\item \hl{it collects spontaneous user messages, based on metadata or specific keywords (\emph{opportunistic sensing})};
    \item \hl{it filters out noise in order to retain only relevant messages, by means of a dedicated deep learning text classifier (\emph{filtering})};
    \item \hl{it selects a subset of users to be contacted in the participatory phase, leveraging a state-of-the-art witness detection classifier based on a rich set of features extracted from text and user metadata (\emph{witness detection})};
    \item \hl{it eventually collects specific information needs from the involved \emph{EOC} personnel;}
    \item \hl{it asks a set of targeted questions to selected users, leveraging benign social media bots, and it collects user replies (\emph{participatory sensing})};
    \item \hl{it filters messages including information about damages to people or infrastructures, by means of a deep learning text classifier (\emph{damage assessment})};
    \item \hl{it enriches messages with geographic information, leveraging a state-of-the-art geoparsing tool, thus tagging messages with the corresponding geographic coordinates (\emph{geoparsing}).}
\end{enumerate}

\hl{The retrieved data can be aggregated and visually summarized by means of information and data visualization techniques, as indicated by the last step of Figure~\mbox{\ref{fig:system}} (\emph{data visualization}).} However, data visualization is out of the scope of this paper. Hence, we just provide crisis mapping as an example of a possible outcome, enabling to highlight those geographic regions that are likely to have suffered higher damage in order to prioritize possible responses~\cite{avvenuti2018crismap,middleton2014real}.

\hl{The previously described steps emphasize a strong interplay between our hybrid sensing strategy and the AI techniques leveraged by our system. In particular, the AI techniques not only provide HERMES with the ability to automatically filter, assess and enrich the contents retrieved from the OSNs, but they directly lead the data gathering during the participatory phase of the hybrid sensing strategy. In fact, they contribute to select which users to contact, targeting them with pertinent questions. Ultimately, the novelty of the HERMES system does not lie in its individual components, but in the peculiar combination of AI and sensing techniques, that has never been proposed and investigated before, to the best of our knowledge. In the remainder, in each separate subsection we provide in-depth description and technical details about each logical component of the system.}

\subsection{Hybrid sensing}
\label{sec:system-hybcrowd}
Concerning data acquisition, HERMES implements an \textit{hybrid sensing} approach, a novel social media-based paradigm that we designed in order to combine the strengths of both \hl{opportunistic and participatory} sensing~\cite{avvenuti2017hybrid}. 

\hl{In a nutshell, the opportunistic sensing phase consists of transparently intercepting data collected and disclosed by users in their unsolicited, spontaneous activities on the OSNs. To do so, the system leverages the filtering features typically provided by the OSN platforms themselves (e.g., keyword-based filtering), and it extracts and enriches the available, relevant information thanks to dedicated, AI-powered techniques. Conversely, the participatory sensing phase involves contacting selected users to solicit more specific and detailed information. Also in this case, dedicated AI techniques enable the system to automatically identify the most valuable users to contact, among those intercepted during the opportunistic sensing phase. Hence, the hybrid sensing approach goes beyond traditional event detection and collection of initial comments -- achievable by opportunistic sensing alone -- up to a direct contact with users, that might have experienced the event firsthand.}

In detail, after having received the notification of an earthquake occurrence from USGS, HERMES initiates the opportunistic phase (\textit{step 1} of Figure~\ref{fig:system}) by crawling one or more social media in order to collect spontaneous, event-related messages, based on specific keywords or metadata. Opportunistic sensing has a twofold goal: 
\begin{enumerate*}[label=(\roman*)]
  \item to figure out, as fast as possible, preliminary situational information, and
  \item to create a list of possible witnesses to solicit for additional, more focused and detailed information during the subsequent participatory phase (\textit{step 5} of Figure~\ref{fig:system}).
\end{enumerate*}
In the participatory sensing phase, benign social media bots automatically ask targeted questions to selected users. Dedicated social media crawlers collect possible answers to these questions. In Figure~\ref{fig:system-seqdiag}, $\Delta_{t}$ represents the latency between the sending of our targeted questions and the possible user replies. Latency has a decisive importance for the applicability of our approach, since enabling timely responses can make the difference during emergencies.

The type of engagement questions sent to witnesses depends on the content of the messages they initially posted and on specific information goals of EOCs. For example, users that sent geolocalized posts are asked about experienced or observed damages to people or infrastructures, while users that posted non-geolocalized posts may be asked to disclose their geographic position.
Henceforth, replies to the former type of questions will be labeled as \textit{reply2damage}, whereas replies to the latter will be labeled as \textit{reply2geo}.
As a politeness best practice, no user is contacted more than once.

It is noteworthy to emphasize that implementing and experimenting with such a hybrid sensing system pose harsh requirements in terms of time, effort and resources, thus jeopardizing the reproduction of experiments. The reason is twofold. First, limitations introduced by the OSNs to regulate the access to their resources (e.g., rate limits imposed on API endpoints) must be faced. Those limits can affect the opportunistic sensing phase, because the huge stream of messages that is typically triggered by mass emergencies may require to deploy multiple crawlers and to refine the filtering strategies in order to match the platform constraints. However, the most severe impact is on the participatory phase, since social bot activities are rigorously restricted in order to prevent spam and deceptive behaviours. Hence, besides carefully selecting possible witnesses to contact, we have to deploy and orchestrate a sufficiently large botnet, ensuring that each bot activity is compliant with the platform policies to avoid possible bans. Second, hybrid sensing is triggered by the occurrence of earthquakes, signaled by the USGS notification service. Moreover, in order to cope with the variety of scenarios that can occur in real applications, we need to test our system on a wide range of earthquake magnitudes and intensities, and diverse epicenter locations, as well as accounting for the variance in urbanization, social media diffusion and usage, and percentage of English speakers.

\subsection{Message filtering and damage assessment}
\label{sec:system-filtering}
\label{sec:system-damage}
As it typically happens in social media analysis, not all the messages collected with specific keywords are actually relevant to the event under investigation. To overcome this issue, we perform message filtering (\textit{step 2} of Figure~\ref{fig:system}) to retain  only relevant messages.
We frame this task as a binary text classification.
Presently, AI state-of-the-art for solving this task is represented by Deep Learning, based on Artificial Neural Networks~\cite{lecun2015deep}. Its main advantages over traditional machine learning can be roughly resumed as: 
\begin{enumerate}
  \item higher performance;
  \item text processed from scratch, with no need of previous feature extraction~\cite{collobert2011natural};
  \item a higher capability to adapt to other problems (i.e., transfer learning) or languages~\cite{semwal2018practitioners}.    
\end{enumerate}
We consider two cascading tasks:
\begin{enumerate}
  \item to filter messages actually related to the monitored crisis event (\emph{relevant messages} hereafter); 
  \item to retrieve messages containing information about the occurrence, or the absence, of damages to people or infrastructures (\emph{damage messages} hereafter).    
\end{enumerate}

\begin{figure*}[t!]
    \centering
    \includegraphics[width=0.7\textwidth]{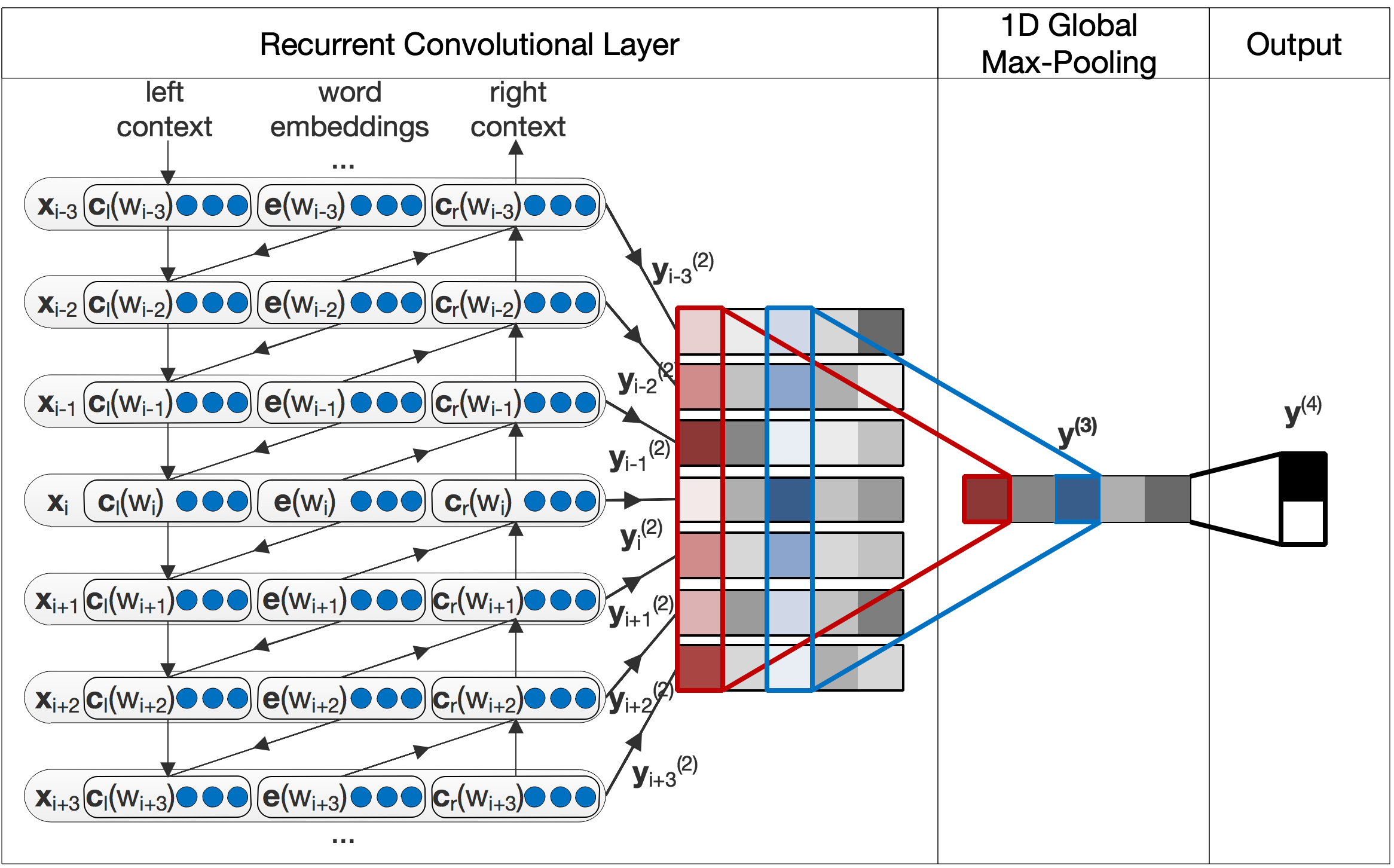}
    \caption{\hl{Architecture of the Recurrent Convolutional Neural Networks used to train the models for message \emph{filtering} and \emph{damage assessment} steps.}}
    \label{fig:rcnn}
\end{figure*}

\hl{To implement \emph{filtering} and \emph{damage assessment} models, we rely on Recurrent Convolutional Neural Networks (RCNNs), combined with max pooling and based on pre-trained word embeddings~\mbox{\cite{mikolov2013distributed}}, adopting the architecture proposed in~\mbox{\cite{lai2015recurrent}}. In~\mbox{\cite{nizzoli2019extremist}}, this architecture was proven to provide particularly good and stable performance under those very challenging conditions, where an unknown but potentially high dataset imbalance is expected. That is, when we want to identify a few pertinent messages hidden in a huge stream of noise. In Figure~\mbox{\ref{fig:rcnn}}, we provide a schema of the architecture. The learning process includes two phases: }
\begin{enumerate*}[label=(\roman*)]
  \item \hl{word representation and}
  \item \hl{text representation learning.}
\end{enumerate*} 
\hl{Word representation learning consists in building, for each word \mbox{$w_{i}$}, a representation of the form}
\begin{equation}
    \label{eq: wrep}
    \mathbf{x}_{i} =  [\mathbf{c}_{l}(w_{i});\mathbf{e}(w_{i});\mathbf{c}_{r}(w_{i})]
\end{equation}

\hl{where \mbox{$\mathbf{c}_{l}(w_{i}), \mathbf{c}_{r}(w_{i})$} are vectors representing the left- and right-side contexts of the word. \mbox{$\mathbf{e}(w_{i})$} is the word representation according to a word embeddings model, that is a continuous, distributional representation of words, pre-trained on large text corpora~\mbox{\cite{mikolov2013distributed}}. The word representation phase is performed by the recurrent convolutional layer, based on a Bidirectional Long-Short Term Memory (BLSTM) architecture, by means of a forward/backward text scan, respectively. The output of this layer are the so called latent semantic vectors \mbox{$\mathbf{y}_{i}^{(2)}$}, obtained by applying a linear transformation together with a \emph{tanh} activation function to \mbox{$\mathbf{x}_{i}$}. 
Text representation learning starts with a 1D global max pooling layer. It is used to convert texts of various lengths into a fixed-sized vector \mbox{$\mathbf{y}^{(3)} = \max_{i=1}^{n} \mathbf{y}_{i}^{(2)}$}, capturing the most important latent semantic factors of the document. The final dense layer obtains the output vector \mbox{$\mathbf{y}_{i}^{(4)}$}, to which a softmax activation function is applied to compute the final output. In order to induce regularization effects and to contrast overfitting, we introduced two dropout layers before and after the recurrent convolutional one, with \mbox{$0.4, 0.3$} dropout probabilities respectively}\footnote{\hl{We refer to~\mbox{\cite{lai2015recurrent}} for further mathematical details and the hyperparameter setting.}}. 

For training and testing our model, we leverage CrisisNLP\footnote{\url{http://crisisnlp.qcri.org/lrec2016/lrec2016.html}}, an open source repository including annotated datasets of tweets related to various types of crises~\cite{imran2016twitter}. \hl{The provided annotations allow to label those tweets according to the binary class \emph{relevant/not relevant}, suitable for the \emph{message filtering} task.} We assemble a single dataset by mixing up all the available, earthquake-related, annotated datasets, and we obtain training ($64\%$), validation ($16\%$) and test ($20\%$) sets by applying a stratified sampling over the target class. The same repository includes also the \emph{Word2Vec} embeddings, trained using crisis-related tweets, that we use in our framework. For the \emph{message filtering} task (\textit{step 2} of Figure~\ref{fig:system}), our model obtains an $AUC = 0.87$ (Area Under the ROC Curve), in line with the state-of-the-art~\cite{nguyen2017robust}. Messages classified as irrelevant are discarded and not used for further analysis. \hl{Hence, this classifier enables HERMES to retain only messages actually pertinent to the monitored earthquake for further analysis.}

The same approach used for filtering messages is also adopted to train a \emph{damage assessment} classifier (\textit{step 6} of Figure~\ref{fig:system}). This time, we label the tweets according to the target class \emph{presence/absence} of damage information.  The classifier is able to detect messages that convey information about the presence of damage to people or infrastructures, reporting a state-of-the-art $AUC = 0.89$. \hl{Thanks to this model, HERMES can provide EOC personnel with fine-grained information about damages, and it can assess the earthquake impact.}

\subsection{Witness detection}
\label{sec:system-witness}
One of the key components of the system (\textit{step 3} of Figure~\ref{fig:system}) is the AI module that selects, among all the users that posted event-related messages, the subset of those to be contacted in the participatory sensing phase. Selecting a subset of users is necessary in order to comply with anti-spamming rules of OSNs, that typically impose upper bounds on the number of messages that can be sent automatically.

Users selection is based on the automatic witness detection system described in~\cite{avvenuti2018realworld}. We cast witness detection as a binary classification task. \hl{The classifier is based on a quadratic Support Vector Machine (SVM), implemented using the LIBSVM framework~\mbox{\cite{chang2011libsvm}}}. The SVM operates on text data and user metadata, retrieved during opportunistic sensing. In particular, it leverages $5$ different classes of features:
\begin{enumerate}\item \hl{linguistic features, representing lexical (e.g., character n-grams, punctuation, etc.) and morpho-syntactic (e.g., Part-of-Speech n-grams) facets of the analysed text;}
  \item word embeddings, \hl{coping with the problem of sparsity of the previous lexical representations of short texts};
  \item sentiment polarity, \hl{tracking the emotional content and the degree of subjectivity
of the text};
  \item semantic entity linking, \hl{enriching the text with semantic references to knowledge graph entities}, and
  \item user metadata, \hl{accounting for message features like the adopted client, the geolocalization, and the posting timestamp}.
\end{enumerate}
\hl{Feature classes 1 and 2 are state-of-the-art, natural language processing approaches to model the text syntax, semantics and writing style. Feature class 3 aims to track subjective and emphasized language frequently used by witnesses. Feature class 4 augments the text semantic representation with rich, structured and inter-linked information. Feature class 5 account for details about the user behaviour, or her temporal and spatial proximity to the event.}

We train and test our witness detector over a semi-automatically labeled dataset of witness tweets~\cite{avvenuti2018realworld}.
Results computed with a 5-fold cross validation report a state-of-the-art $F1 = 0.85$, obtained for messages classified with high machine confidence scores~\cite{avvenuti2018realworld}. \hl{The obtained witness detection model enables HERMES to select the most promising users to contact during the participatory sensing phase. In this way, the system can obtain relevant, first-hand information while coping with the severe OSN platform limitations on bot activities.  }

\subsection{Geoparsing}
\label{sec:system-geoparsing}
Geospatial analysis suffers from the fact that only a small fraction of OSN messages are natively geotagged~\cite{avvenuti2018gsp}.
For instance, the ratio of geotagged Twitter messages is in the range of $1\%$ to $4\%$~\cite{avvenuti2018gsp}, which indeed constitutes a severe limitation.
To increase the number of geotagged messages, we employ our geoparsing module (\textit{step 7} of Figure~\ref{fig:system}), based on Geo-Semantic-Parsing (GSP). As described in~\cite{avvenuti2018gsp}, GSP receives a text document as its input and returns an enriched document where all mentions of places are associated to the corresponding geographic coordinates. 

First, semantic annotation is performed to identify relevant parts of the input text (\emph{anchors}) that can be linked to pertinent resources (e.g., DBpedia entities) in the Linked Data cloud. \hl{We perform semantic annotation leveraging the state-of-the-art, off-the-shelf tool \emph{Tagme}~\mbox{\cite{ferragina2011fast}}, specifically designed to work with short, poorly written texts, and thus being particularly suitable for social media applications. Semantic annotation represents the key novelty of the GSP approach over the state-of-the-art benchmarks. Its main advantage is represented by the ability of the semantic annotators to leverage the information-rich, interlinked structure of the knowledge graphs to implicitly perform disambiguation. In fact, toponymic polysemy poses the main limitations to traditional approaches, based on named-entity recognition and geographic gazetteer lookup. }

\hl{Then, the GSP tool expands the available information by traversing the Linked Data, leveraging the Resource Description Framework (RDF) properties. In particular, it retrieves resources semantically equivalent to those returned by \emph{Tagme}, but belonging to other knowledge graphs. At this point, a specific machine learning classifier, based on a Support Vector Machine, prunes those resources that do not correspond to geographic locations. It leverages several, carefully crafted features, to measure the pertinence between the anchors and the corresponding semantic resources, and to assess the likelihood that the semantic resources actually refer to geographic locations. To name just a few, these features include the edit distance between the anchors and the semantic resource titles, the Part-of-Speech tags of the anchors, the ontology classes of the semantic resources, and the number of occurrences of the anchors in the resource abstracts~\mbox{\cite{avvenuti2018gsp}}.  Finally, the GSP tool parses the relevant resources to extract the geographic coordinates, thus geotagging the related anchors.}

GSP has a number of advantages over other geoparsing approaches: 
\begin{enumerate*}[label=(\roman*)]
  \item it does not require any explicit geographic information (e.g., GPS coordinates, location information, time zones);
  \item it only exploits text data of input documents (e.g., it does not require any user information or social network topology);
  \item it processes one text document at a time (e.g., it does not require all tweets from a user’s timeline, or many documents on  a  given  topic);
  \item it  does  not  require  users  to  specify  a target geographic region but, instead, it geoparses and geotags places all over the world;
  \item by leveraging Linked Data, GSP is capable of extracting fine-grained, structured geographic information (e.g., country $\rightarrow$ region $\rightarrow$ city $\rightarrow$ building).
\end{enumerate*}

\hl{In~\mbox{\cite{avvenuti2018gsp}}, GSP was proved to largely outperform state-of-the-art benchmarks, reaching performance suitable for realistic applications. Hence, GSP provides HERMES with the ability to identify and geotag toponyms mentioned by OSN users, thus enabling useful applications like crisis mapping. }

\subsection{Legal and ethical considerations on the Hybrid Sensing approach}
\label{subsec:ethics}
\hl{The hybrid sensing approach of the HERMES system raises legal and ethical considerations, that always come with social crisis data gathering. In fact, this effort is deployed when people are at their most vulnerable, disclosing personal information on OSNs to seek help, update or trace back loved ones\mbox{~\cite{crawford2015limits}}.} 

\hl{From a legal point of view, the experimental campaings conducted with HERMES on the Twitter platform accomplish the legal requirement of the European General Data Protection Regulation (GDPR) in terms of personal data processing, and the Twitter platform terms of use. In particular, the platform cleary informs users that ``Twitter is public and Tweets are immediately viewable and searchable by anyone around the world.''}\footnote{https://twitter.com/en/privacy}\hl{. Hence, provided that a proper treatment is guaranteed, data can be crawled without an explicit user consent. Moreover, during the participatory phase, the usage of bots and the objectives of the campaign were made clear by the approach tweets and the automated account profile information, and the solicited users consciously accepted to disclose personal data.}

\hl{However, compliance with the law does not solve ethical concerns \emph{per se}, and the issue of informed consent is not the end of the story. In fact, ``even well informed and rational individuals cannot
appropriately self-manage their privacy'', given the ability of big data approaches to aggregate multiple data sources to reveal more information~\mbox{\cite{solove126privacy}}. This is an open issue in social crisis data research~\mbox{\cite{crawford2015limits}}. On the one hand, the particular context of crisis may induce to reconsider priorities, making acceptable some privacy leakage in exchange of a more effective and prompt response to the emergency, which includes also contributing to a research effort aiming to provide better emergency management tools for future events. On the other hand, a condition of higher people vulnerability should induce more effort in their privacy protection. Solving this Hamletic dilemma is beyond the scope of this work. Our commitment was entirely devoted to guarantee a proper, secure and anonymized data treatment. }

%
 \section{Real-world experiment and results}
\label{sec:results}

\begin{table*}[t]
\centering
\begin{adjustbox}{max width=\textwidth}
\begin{tabular}{@{}l r r c r r r r r r r r @{}}
\toprule
\multicolumn{3}{c}{} & \textbf{date} & \textbf{collected} & \textbf{relevant} & \textbf{collected} & \textbf{collaborative} & \textbf{message} & \textbf{reply latency $\Delta_t$} & \hl{\textbf{participants wrt }} \\
\textbf{place}           & \textbf{magnitude} & \textbf{depth (km)} & \textbf{(2015)} & \textbf{messages} & \textbf{messages} & \textbf{replies} & \textbf{replies} & \textbf{gain} & \textbf{(minutes)} & \hl{\textbf{USGS-DYFI}} \\
\\ \midrule
San Ramon, California    & 3.5 & 11 & 02/04 & 2,266    & 836   & 164 & 78\% & $+$20\%& 5 & \hl{4\%} \\
Lila, Philippines        & 4.8 & 81 & 30/03 & 2,396    & 868   & 161 & 95\% & $+$19\%& 11 & \hl{215\%} \\
Lamjung, Nepal           & 7.5 & 12 & 25/04 & 117,774  & 8,545 & 160 & 95\% & $+$2\% & 24 & \hl{15\%} \\
Kokopo, Papua New Guinea & 7.7 & 66 & 29/03 & 10,576   & 672   & 153 & 96\% & $+$23\%& 28 & \hl{1,958\%} \\
Irving, Texas            & 3.3 & 6  & 02/04 & 2,044    & 620   & 132 & 87\% & $+$21\%& 8 & \hl{16\%} \\                     
\\ \bottomrule
\end{tabular}
\end{adjustbox}
\caption{Statistics about the top-5 earthquakes monitored by HERMES, \hl{and a comparison between the amount of users engaged by our system and those contributing to the USGS-DYFI service for the same events.}}
\label{tab:stats_quakes}
\end{table*} 
In this section, we present a thorough evaluation of HERMES, with the aim of assessing the effectiveness of our approach in increasing the amount of social crisis data and enhancing the information conveyed. In particular, we evaluate the intake of the participatory sensing phase, by comparing the amount and quality of the solicited replies with respect to the opportunistically crawled messages. As already highlighted in Section \ref{sec:system-hybcrowd}, testing such a system is a complex task, as it requires the capability of promptly involving real users during the unpredictable occurrence of earthquakes of medium-to-high magnitude, while coping with the limitations imposed by the OSN platforms.

\subsection{Increasing the Amount of Social Crisis Data}
\label{sec:general-results}

We tested HERMES in-the-wild on Twitter by running a long-lived experiment lasting from February to May, 2015. During the experiment, our system was triggered by USGS to monitor up to 436 earthquakes. 
In order to conveniently report the experimental results, we introduce the following definitions:
\begin{itemize}
  \item \emph{collected messages}: the tweets gathered through opportunistic sensing;
  \item \emph{relevant messages}: the tweets actually concerning the monitored earthquake, retained (e.g., not discarded) by the filtering module described in Section~\ref{sec:system-filtering};
  \item \emph{collected replies}: the messages received from users contacted in the participatory phase;
  \item \emph{collaborative replies}: the percentage of collected replies containing valuable information;
  \item \emph{message gain}: the ratio of collected replies to relevant messages;
  \item \emph{reply delay} $\Delta_{t}$: the average latency of replies.
\end{itemize}
Table~\ref{tab:stats_quakes} shows the results from the top-5 earthquakes in terms of the number of collected replies, according to the  defined measures. These earthquakes occurred in different geographic areas, featuring diverse characteristics in terms of demography, economy, urbanization and language. Moving the focus on the earthquake intensity, the examined range spans from moderate events (California, Texas) to fatal disasters involving thousands of casualties (Nepal).

The participatory component of HERMES was able to produce a remarkable message gain of about $20\%$. 
One notable exception was the devastating Nepal earthquake, with a message gain of only $2\%$ despite a number of replies comparable to the other events. This result was mainly due to the limited amount of targeted questions allowed by the Twitter platform during the participatory phase, compared to the large number of relevant messages collected during the opportunistic phase. In any case, we envision the possibility to obtain similar message gains also for major events by deploying larger botnets to cope with the platform limitations.

\hl{Notably, most of the collected replies conveyed valuable information (collaborative replies). In detail, collaborative replies range from 78$\%$ for the little California earthquake to more than 95$\%$ for all the three intense ones. These results confirm a general tendency of users to collaborate, which appears to increase with the severity of the earthquake. In Section~\mbox{\ref{subsec:ethics}}, we emphasised that the user propensity to disclose personal information depends on the context. In fact, severe earthquakes may induce users to accept potential privacy leakages (like those related to position disclosure) in exchange of a more effective and prompt response to their needs during the emergency~\mbox{\cite{crawford2015limits}}. Results reported in Table~\mbox{\ref{tab:stats_quakes}} confirm this intuition.}

The \hl{second-last} column of Table~\ref{tab:stats_quakes} confirms that participatory sensing retrieved new information in a timely fashion, with a reply latency $\Delta_t \leq 28$ minutes. Similar results were already observed in~\cite{avvenuti2017hybrid}, and they accomplish the need of a prompt response to disasters.

\hl{Finally, in the last column of Table~\mbox{\ref{tab:stats_quakes}} we compare the amount of HERMES collaborative participants with respect to those contributing to the USGS \emph{Did You Feel It?} (DYFI)}\footnote{\url{earthquake.usgs.gov/data/dyfi/}} \hl{web service, for the same earthquake events. USGS-DYFI requires users to spontaneously navigate to the corresponding web page and fill a questionnaire about their earthquake experience. Results show that HERMES, despite being a newcomer, was able to engage far more users than a well-established, institutional service like USGS-DYFI, when considering events occurring outside the U.S. and with poor media coverage (Philippines and Papua New Guinea). Interestingly, in~\mbox{\cite{mak2016makes}} authors demonstrated that the USGS-DYFI ability to engage participants strongly depends on both the physical features of the earthquake (e.g., magnitude, depth) and the demographic traits of the involved area. For example, the response rate increases with the percentage of English native speakers, the education and the wealth of the population, whereas it decreases with the percentage of immigrants and the median age. Hence, the capability of HERMES to directly contact witnesses proves crucial to engage users that are not aware of services like USGS-DYFI. Moreover, people reluctance to participate in surveys -- especially long and complex ones -- is well known in several domains~\mbox{\cite{sahlqvist2011effect, jepson2005mailed, parashos2005response}}. This tendency is further confirmed by another experimental campaign that we conducted in autumn 2016, when we monitored 626 earthquakes with magnitude $>$2.5 following a hybrid sensing approach, similar to the one adopted in HERMES. However, in this case we targeted users with tweets inviting to follow a link to a simple Web survey questionnaire. In the face of 2,384 approach tweets, we engaged only 75 participants, corresponding to an unsatisfactory 3.1\% response rate. As a consequence, the capability of HERMES to handle the whole interaction with users on the same OSN platform appears as a key ingredient in order to engage less helpful or trustful users, not willing to browse an external Web page to participate in a survey.}

\hl{Since now, we described how HERMES can increase the amount of social crisis data, operating in an orthogonal and complementary way with respect to services like USGS-DYFI, especially during emergencies affecting developing countries or having poor media coverage. In the remainder of this section, we assess the amount and quality of the information conveyed by the retrieved data.  }

\subsection{Enhancing the information conveyed by Social Crisis Data}
\label{sec:case-study-results}

In the previous Section, we demonstrated that the participatory sensing ensures a gain of about $+20\%$ in the amount of available messages. Now, we focus on the content of the messages collected during the five earthquakes listed in Table~\ref{tab:stats_quakes}. 

\begin{figure*}[t]
    \begin{subfigure}[t]{0.30\textwidth}
        \includegraphics[width=\textwidth]{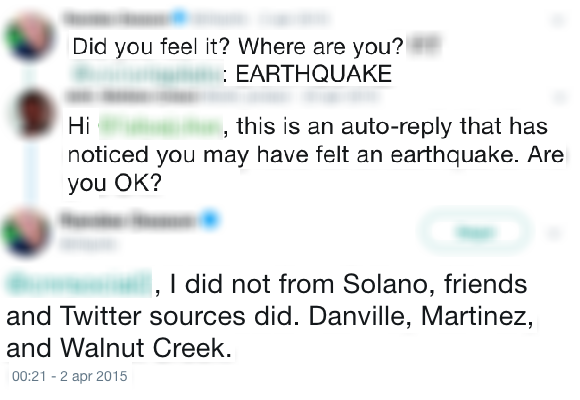}
        \caption{} 
        \label{subfig:1wasnt_there}
    \end{subfigure}\hspace{0.04\textwidth}\begin{subfigure}[t]{0.30\textwidth}
        \includegraphics[width=\textwidth]{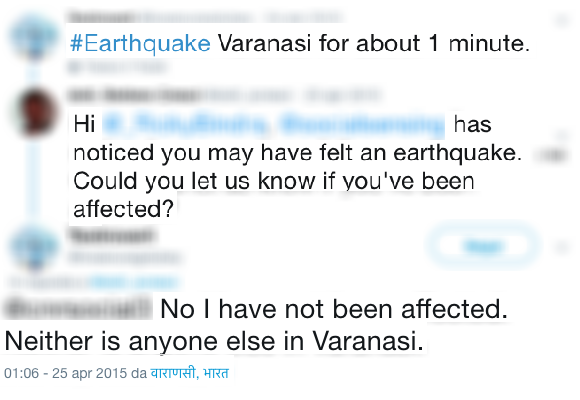}
        \caption{}
        \label{subfig:2notaffected}
    \end{subfigure}\hspace{0.04\textwidth}\begin{subfigure}[t]{0.30\textwidth}
        \includegraphics[width=\textwidth]{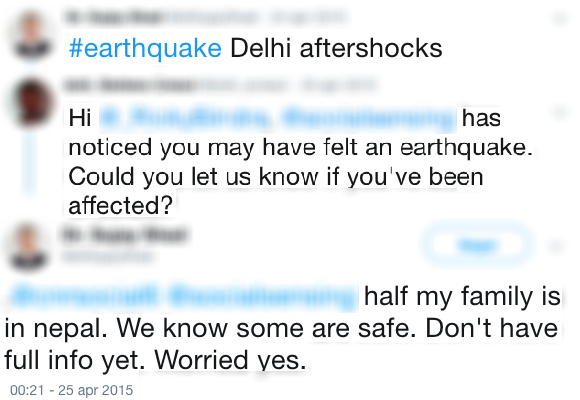}
        \caption{} 
        \label{subfig:3worried_about_family}
    \end{subfigure}\\
    \begin{subfigure}[t]{0.30\textwidth}
        \includegraphics[width=\textwidth]{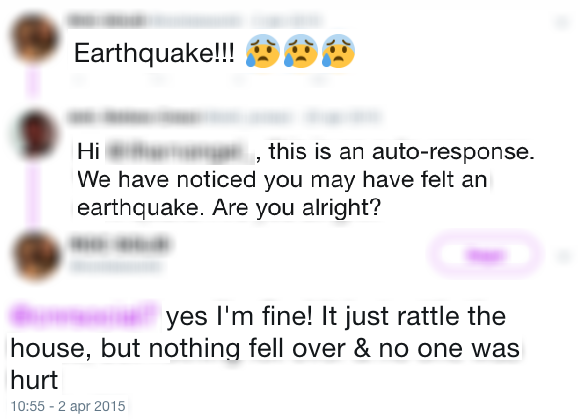}
        \caption{}
        \label{subfig:10felt_no_harm}
    \end{subfigure}\hspace{0.04\textwidth}\begin{subfigure}[t]{0.30\textwidth}
        \includegraphics[width=\textwidth]{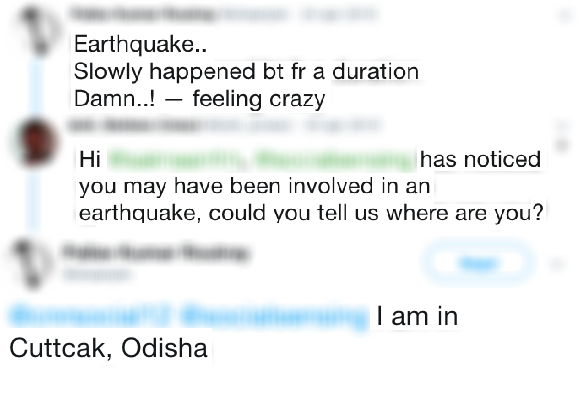}
        \caption{}
        \label{subfig:12directly_where}
    \end{subfigure}\hspace{0.04\textwidth}\begin{subfigure}[t]{0.30\textwidth}
        \includegraphics[width=\textwidth]{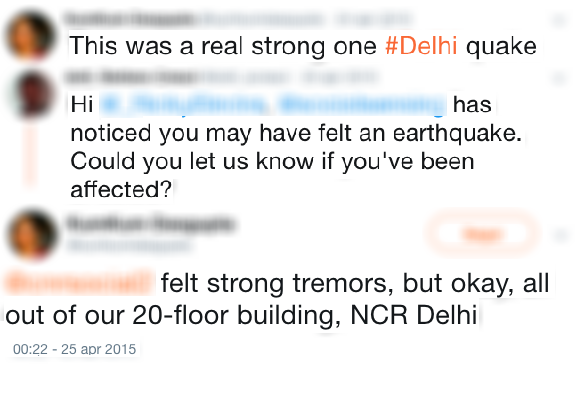}
        \caption{} 
        \label{subfig:5felt_and_panic}
    \end{subfigure}\\
    \begin{subfigure}[t]{0.29\textwidth}
        \includegraphics[width=\textwidth]{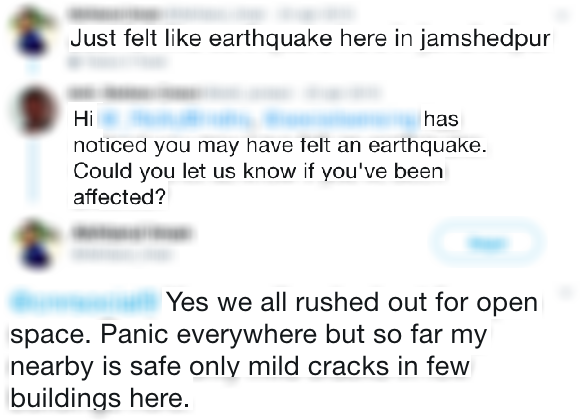}
        \caption{}
        \label{subfig:4panic}
    \end{subfigure}\hspace{0.04\textwidth}\begin{subfigure}[t]{0.30\textwidth}
        \includegraphics[width=\textwidth]{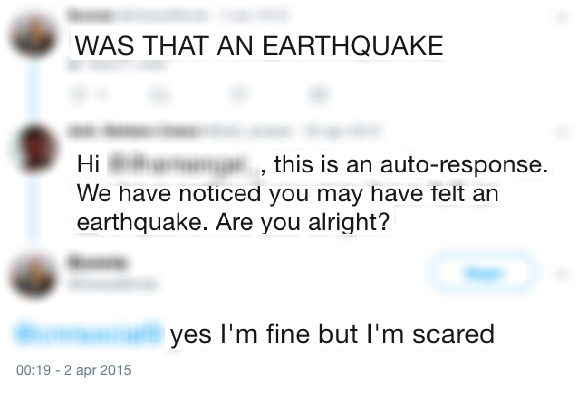}
        \caption{}
        \label{subfig:6fine_but_scared}
    \end{subfigure}\hspace{0.04\textwidth}\begin{subfigure}[t]{0.30\textwidth}
        \includegraphics[width=\textwidth]{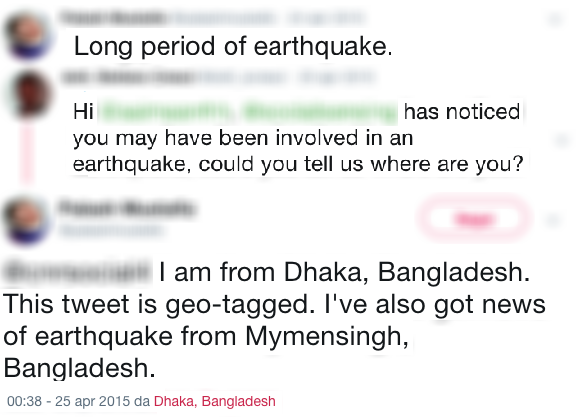}
        \caption{} 
        \label{subfig:7after_geotagged}
    \end{subfigure}\caption{Examples of ``conversations'' with Twitter users. Conversations are composed of 3 tweets: (i) the topmost tweet is a spontaneous message collected with opportunistic sensing; (ii) the middle tweet is the targeted question automatically sent by our system, and (iii) the bottom tweet is the user reply to our question.}\label{fig:overall-results}
\end{figure*}

Figures~\ref{subfig:1wasnt_there}--\ref{subfig:12directly_where} show some examples of the user replies to our targeted questions. In all the provided examples, the replies solicited in the participatory phase integrated, increased or clarified the information conveyed by the original, spontaneous tweet. In Figure~\ref{subfig:1wasnt_there}, the user spontaneously signaled an earthquake, asking who felt it. From the reply to our question, concerning her safety, we learned that the earthquake was not perceivable in the user location, but it was felt nearby. The user in Figure~\ref{subfig:12directly_where} reported an earthquake that she clearly felt without suffering consequences, but it was thanks to the solicited reply that we could locate her. The original tweet of Figure~\ref{subfig:2notaffected} signaled a one minute-long jolt affecting a large city, but we could exclude serious consequences after the interaction with the user. While replies in Figures \ref{subfig:10felt_no_harm} and \ref{subfig:6fine_but_scared} attested different reactions from people involved in moderate events without serious consequences, the ones in figures~\ref{subfig:5felt_and_panic} and \ref{subfig:4panic} told about people run out of the buildings, with the last one reporting also damages to the structures. 

The propensity of people involved in severe events to actively collaborate with us is shown in Figures~\ref{subfig:3worried_about_family} and \ref{subfig:7after_geotagged}. In the first one, we show the distressing testimony of a person tweeting from Delhi, who was worried about some relatives close to the Nepal earthquake epicenter and not yet traced. In a real application, this kind of witness may be put in direct contact with a first responder with mutual benefit. In the last one, we asked the position to a user reporting an earthquake. She did not limit herself to reply specifying her location, but she had also the foresight of enabling the Twitter GPS geotagging feature for providing her exact coordinates.      

After this qualitative exploration of the collected replies, we proceed to their quantitative analysis. We want to compare solicited replies to spontaneous messages in terms of the density, variety and granularity of the conveyed information. We focus on damage assessment and geolocation, leveraging the dedicated modules described in sections~\ref{sec:system-damage},~\ref{sec:system-geoparsing}.   

\subsubsection{Damage assessment} 

\begin{figure}[ht!]
    \centering
    \includegraphics[width=0.6\textwidth]{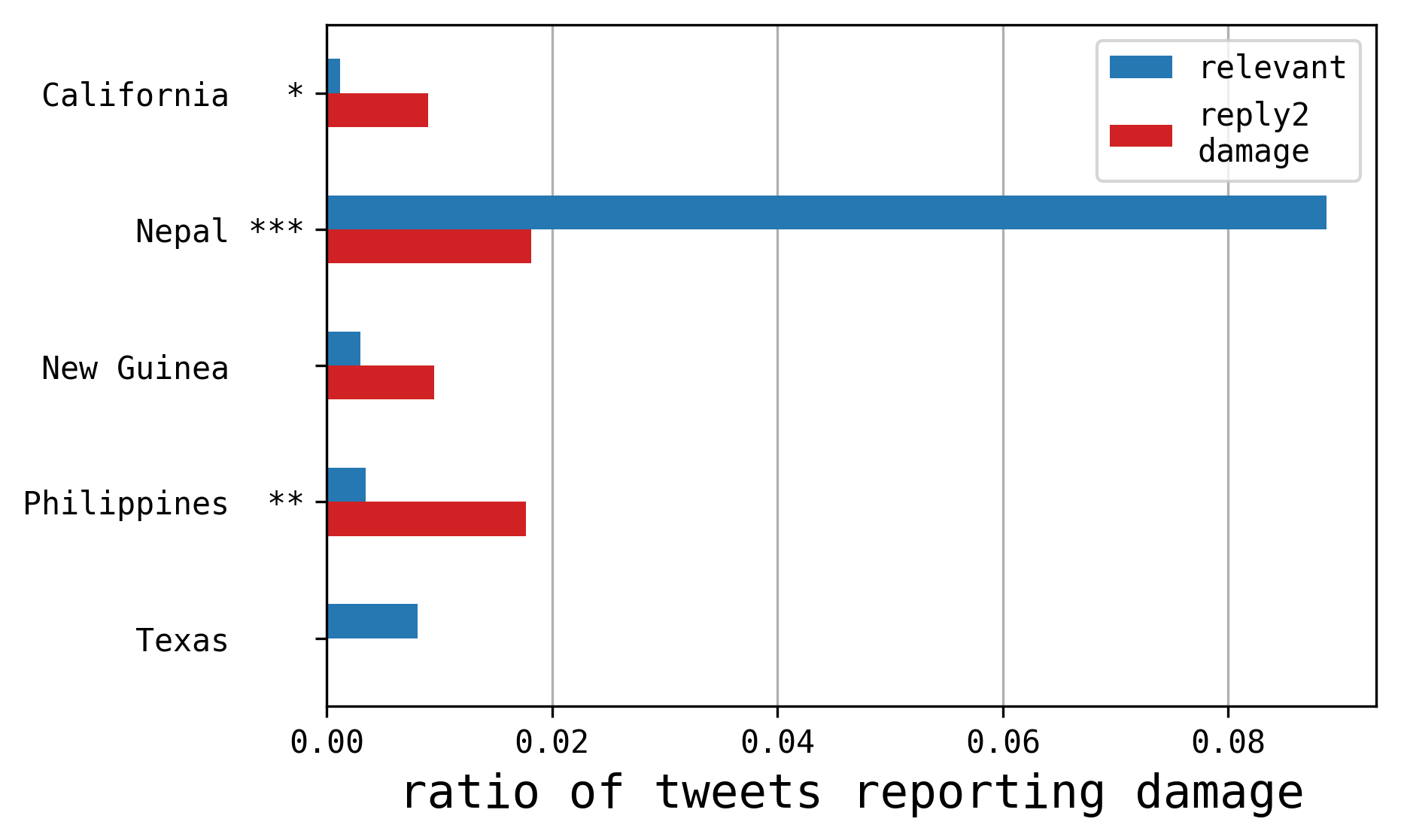}
    \caption{Ratio of tweets reporting the presence of damage in solicited replies (\emph{reply2damage}) with respect to spontaneous \emph{relevant} tweets. Statistical significance of comparisons is evaluated by means of T-tests, with \texttt{***}: $p < 0.01$; \texttt{**}: $p < 0.05$; \texttt{*}: $p < 0.1$.}
    \label{fig:damage}
\end{figure}

HERMES enables a near real-time damage assessment in the aftermath of crisis events, by leveraging messages reporting the presence of damage (or lack thereof) to people or infrastructures. Those messages are conveniently filtered by the dedicated module described in Section~\ref{sec:system-damage}, based on a state-of-the-art, deep learning text classifier.

We evaluated the system performance in retrieving damage-related information during the five earthquakes listed in Table~\ref{tab:stats_quakes}. Figure~\ref{fig:damage} shows the ratio of tweets reporting damage information, distinguishing spontaneous tweets (\emph{relevant}) with respect to solicited replies to questions about the presence of damages (\emph{reply2damage}). 
For harmless earthquakes, there was a low percentage of damage-reporting tweets in spontaneous (\emph{relevant}) messages. Notably, the percentage significantly increased (up to $7\times$) when considering replies to our targeted questions (\emph{reply2damage}). 
Focusing on the destructive Nepal event, we found a relevant $9\%$ of damage-reporting tweets in \emph{relevant} messages, whereas for \emph{reply2damage} messages the fraction was less than $2\%$, in line with other events. This can be explained by the fact that the majority of users was tweeting from surrounding countries (e.g., India) rather than from the epicenter area (cfr. figures~\ref{subfig:2notaffected},~\ref{subfig:3worried_about_family},~\ref{subfig:12directly_where},~\ref{subfig:5felt_and_panic},~\ref{subfig:4panic},~\ref{subfig:7after_geotagged}). This might be due to the impossibility to communicate from the devastated area, where communication infrastructures suffered severe damages. Hence, the majority of replying witnesses reported to have felt the shake from India and Bangladesh, without suffering any damage. Instead, spontaneous tweets reporting damages might be based on second-hand information.

\begin{figure}[t]
    \centering
    \includegraphics[width=0.6\textwidth]{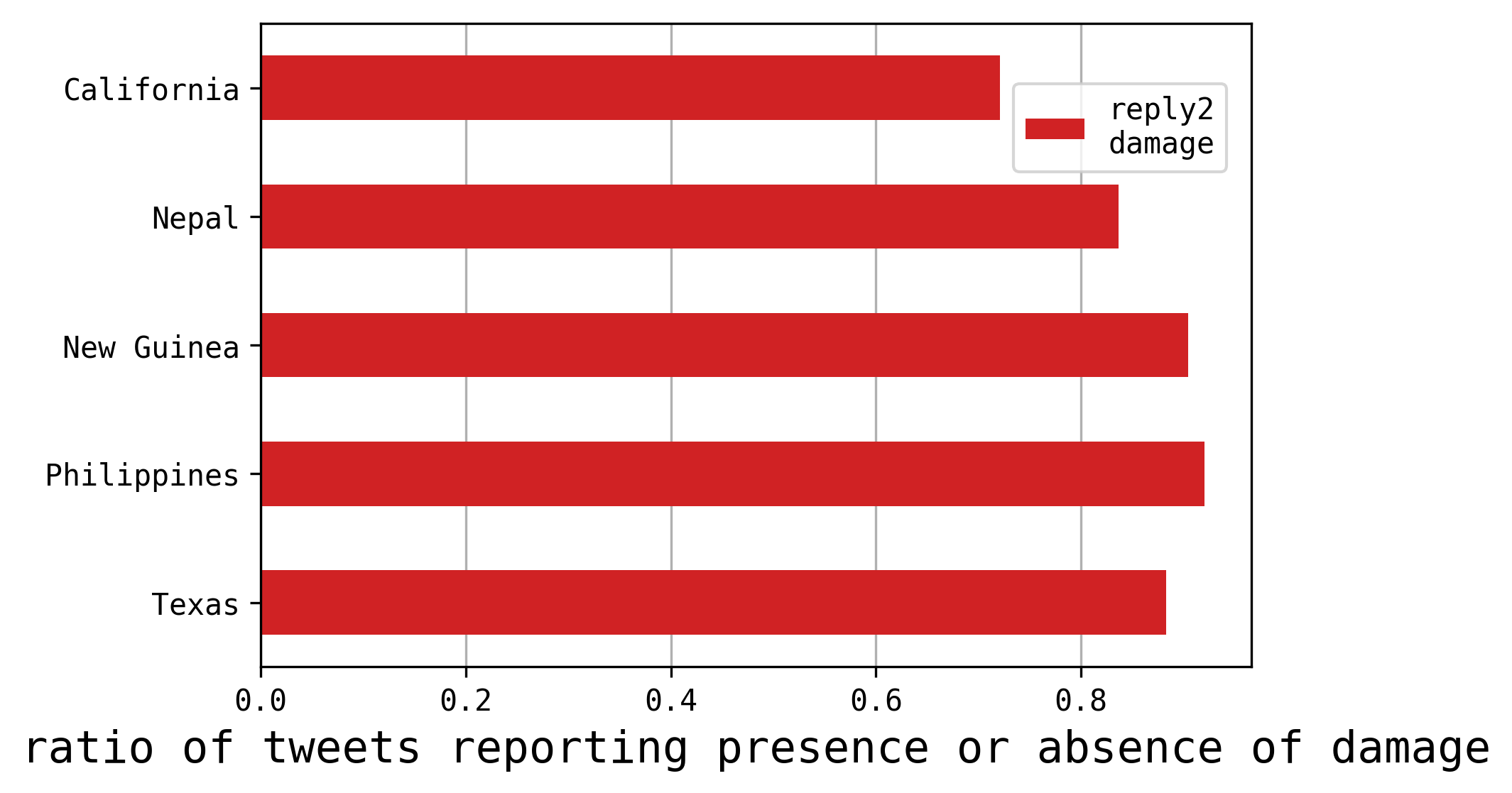}
    \caption{Ratio of tweets reporting information about damage (presence or absence) in solicited replies (\emph{reply2damage}). }
    \label{fig:damage_man}
\end{figure}

In order to have a better insight on the information conveyed by the participatory sensing, we also considered the \emph{reply2damage} messages reporting the lack of damage. Results are shown in figure~\ref{fig:damage_man}. Solicited replies proved to be very effective in real-time confirmation of the absence of damages, with more than $80\%$ conveying this information. This is a non-trivial, significant contribution in the aftermath of a crisis event. In a scenario similar to the one occurred in the Nepal earthquake, by combining information about non-damaged areas to the presence of non-communicating ones, responders can figure out where to focus their attention. 

The high collaboration rate in user replies encourages to devote further experiments to enhance the performance of the participatory component of the hybrid sensing in detecting the presence of damage, by enriching the set of questions asked.

\begin{figure}[t]
    \centering
    \includegraphics[width=0.6\textwidth]{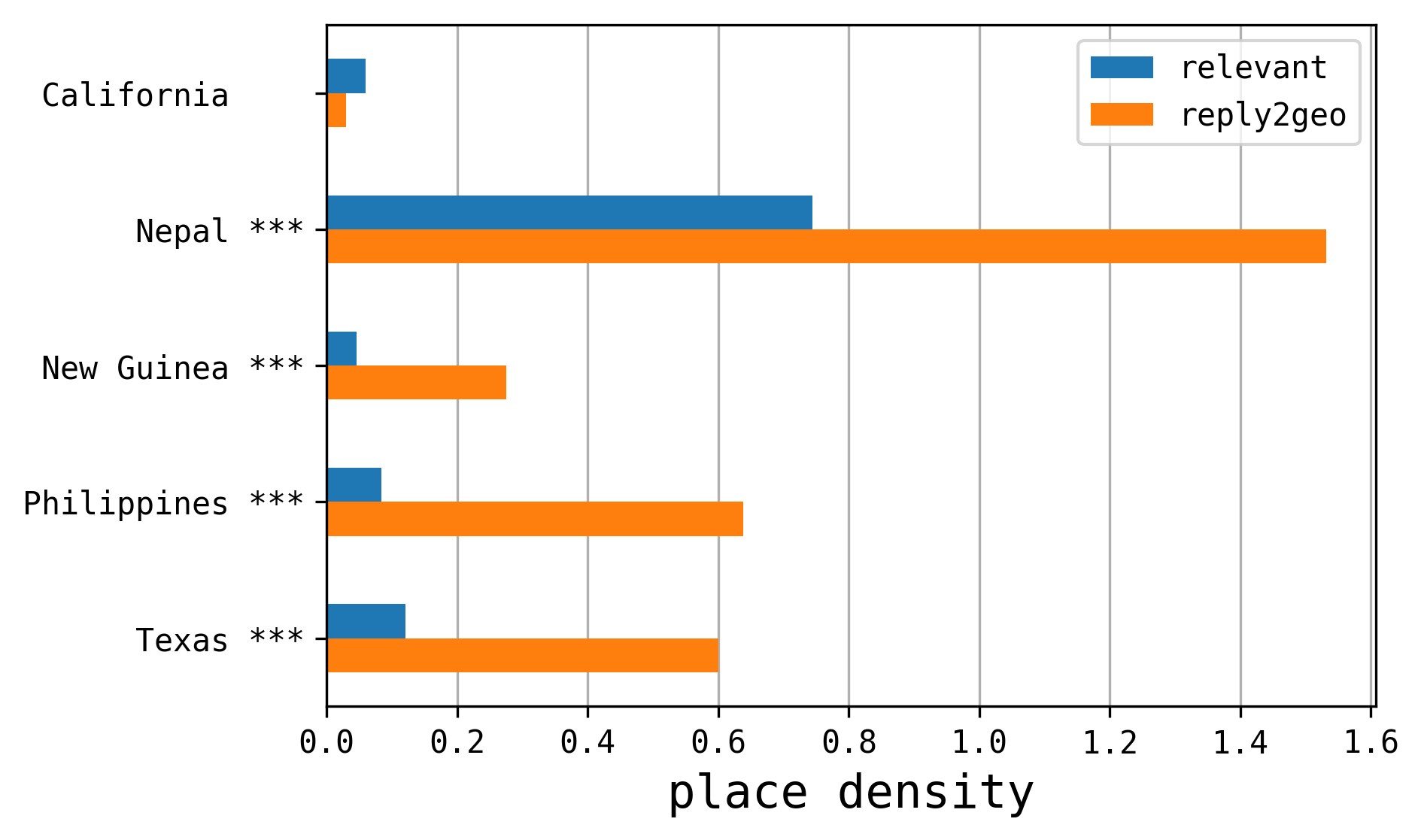}
    \caption{Place density per event for solicited replies (\emph{reply2geo}) and spontaneous \emph{relevant} tweets. Statistical significance of comparisons is evaluated by means of T-tests, with \texttt{***}: $p < 0.01$; \texttt{**}: $p < 0.05$; \texttt{*}: $p < 0.1$.}
    \label{fig:place-norm}
\end{figure}

\subsubsection{Geolocation} 

The other key ingredient to convert generic information to actionable knowledge is geolocation. HERMES leverages the geoparsing module, described in Section~\ref{sec:system-geoparsing}, to automatically extract geographic references from text and tag them with the corresponding coordinates. As in the previous Section, we are interested in comparing the information retrieved through traditional opportunistic sensing with the contribution added by the participatory phase of our hybrid approach. For this purpose, we define the following measures, accounting for diverse facets related to geographic information:
\begin{itemize}
    \item \emph{place density}: the average number of places mentioned per message;
    \item \emph{place variety}: the average number of \emph{distinct} places mentioned per message;
    \item \emph{coverage gain}: ratio of new places discovered in solicited replies (\emph{reply2geo}) to places already known from spontaneous \emph{relevant} tweets.
    \item \emph{place granularity distribution}: percentage of places belonging to each geographic granularity level. We consider the following granularity levels, in order of decreasing specificity: \emph{building}, \emph{city}, \emph{region}, \emph{country}, \emph{other}.  
\end{itemize}

\begin{figure}[t]
    \centering
    \includegraphics[width=0.6\textwidth]{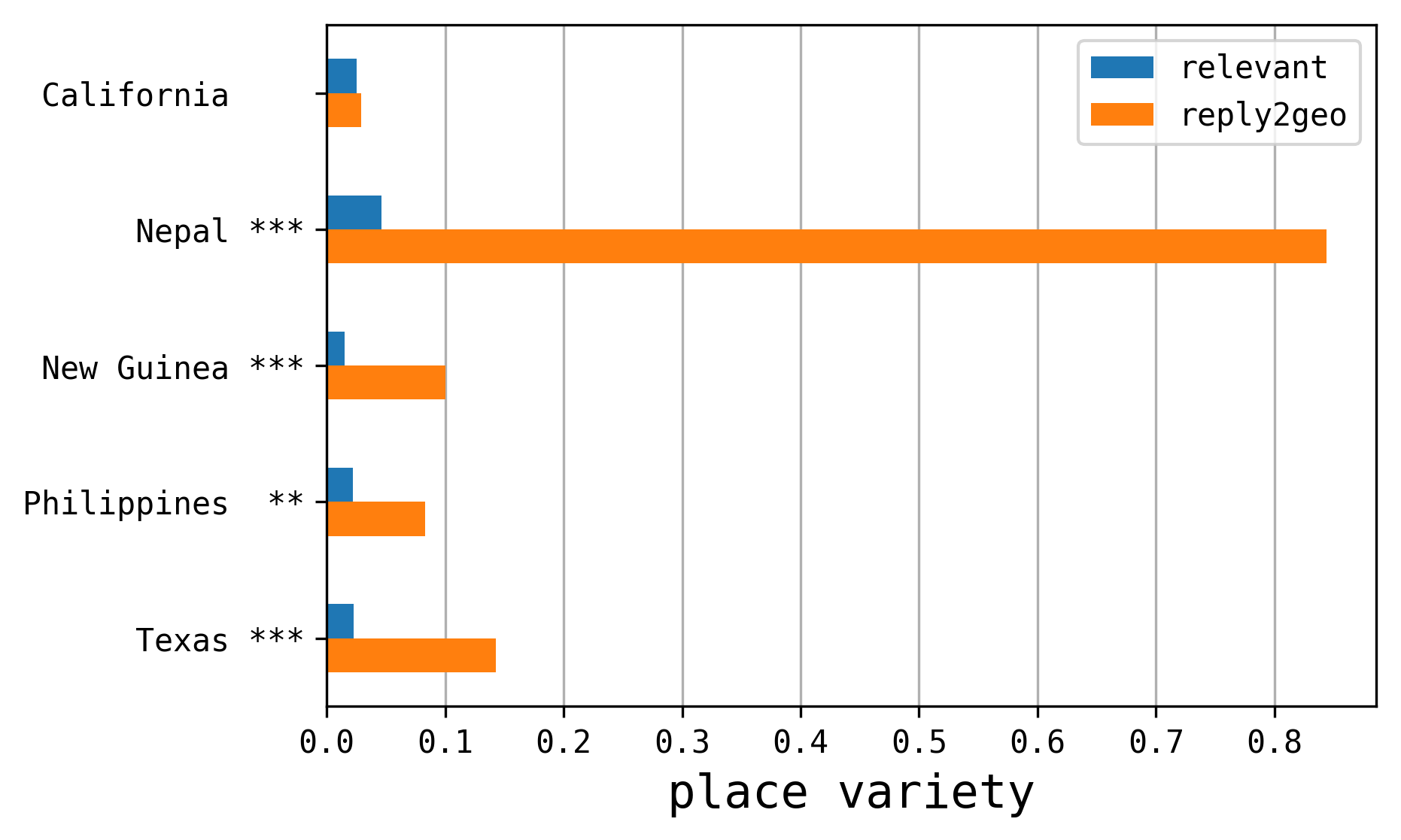}
    \caption{Place variety per event for solicited replies (\emph{reply2geo}) and spontaneous \emph{relevant} tweets. Statistical significance of comparisons is evaluated by means of T-tests, with \texttt{***}: $p < 0.01$; \texttt{**}: $p < 0.05$; \texttt{*}: $p < 0.1$.}
    \label{fig:dist-place-norm}
\end{figure}

\begin{figure}[t]
    \centering
    \includegraphics[width=0.6\textwidth]{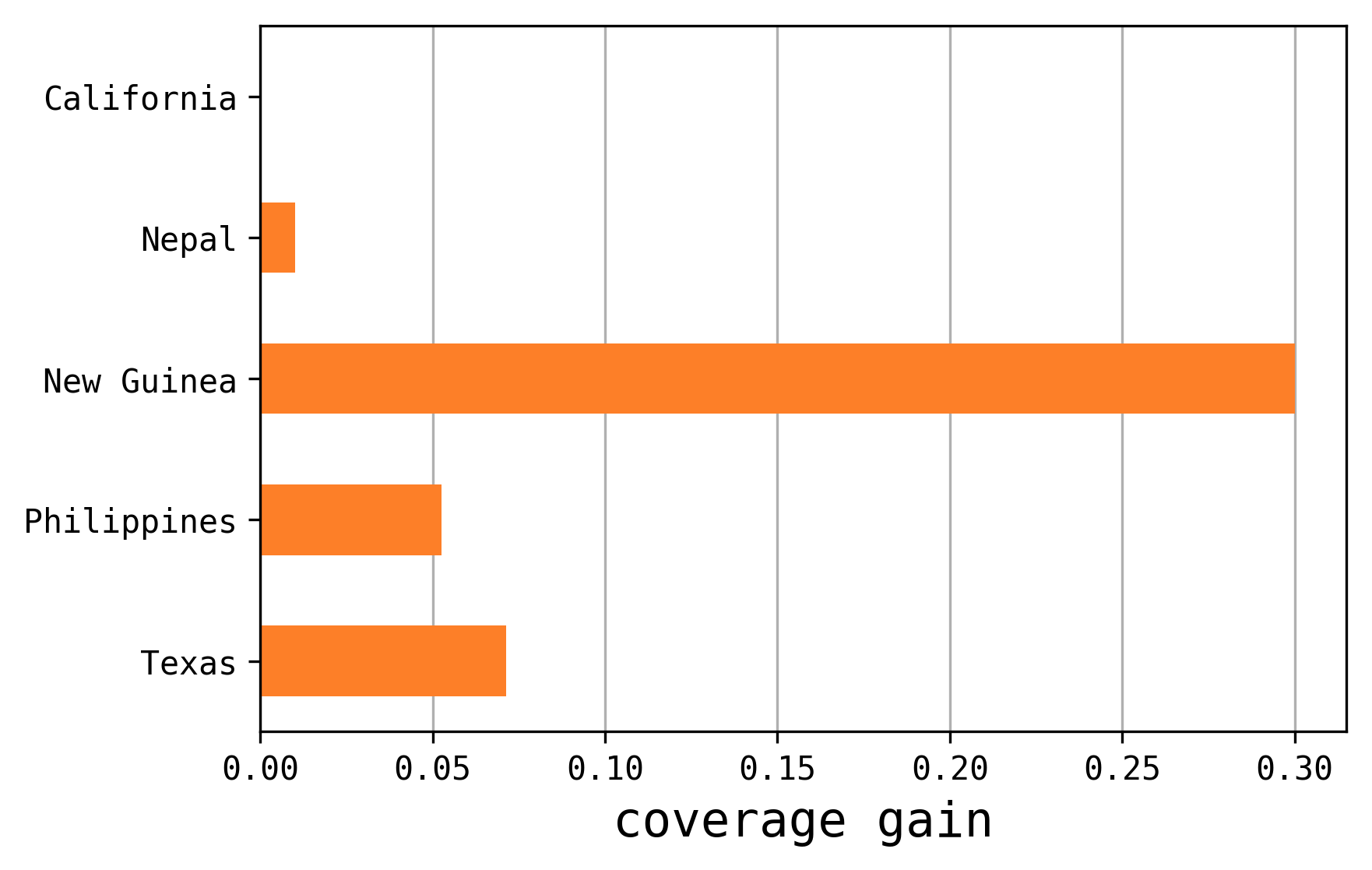}
    \caption{Coverage gain obtained through solicited replies (\emph{reply2geo}) with respect to spontaneous \emph{relevant} tweets.}
    \label{fig:new-place-norm}
\end{figure}

\begin{figure}[h]
    \centering
    \includegraphics[width=0.6\textwidth]{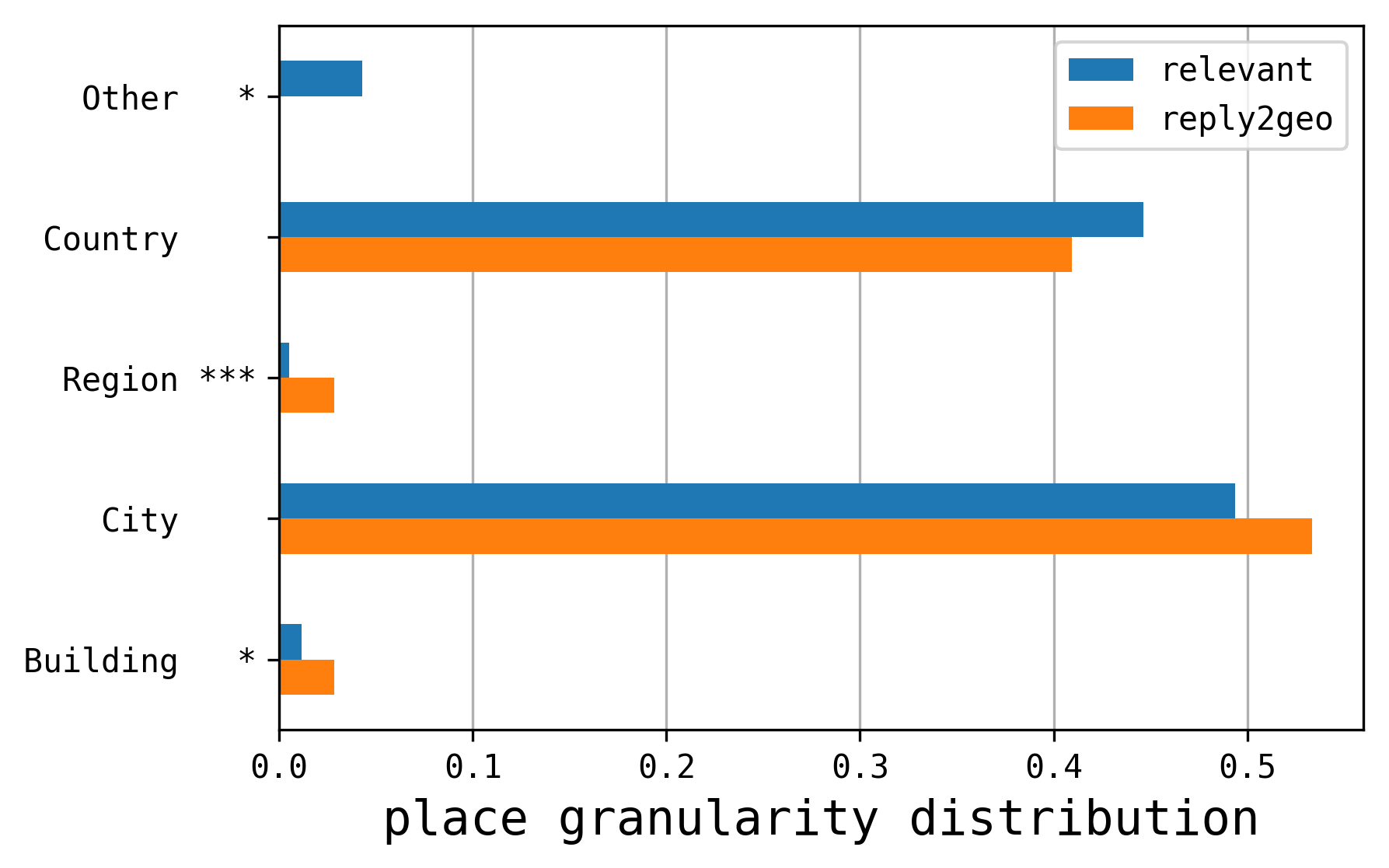}
    \caption{Place granularity distribution in solicited replies (\emph{reply2geo}) with respect to spontaneous \emph{relevant} tweets. Statistical significance of comparisons is evaluated by means of T-tests, with \texttt{***}: $p < 0.01$; \texttt{**}: $p < 0.05$; \texttt{*}: $p < 0.1$.}
    \label{fig:place-types}
\end{figure}

Figures~\ref{fig:place-norm}-\ref{fig:place-types} aggregate the results obtained during the five earthquakes listed in Table~\ref{tab:stats_quakes}, according to the metrics introduced. In detail, Figure~\ref{fig:place-norm} reports the \emph{place density} per event, distinguishing between solicited replies (\emph{reply2geo}) and spontaneous \emph{relevant} messages. Solicited replies had a significantly larger density of places with respect to spontaneous \emph{relevant} messages. Their ratios spanned from $2\times$ for Nepal, up to more than $7\times$ for  Philippines. The only exception was California, which anyway was not statistically significant. \hl{Notably, these results further confirm that users involved in more severe earthquakes have higher propensity to disclose personal information, despite possible privacy leakages, as already discussed in Section~\mbox{\ref{sec:general-results}}}. Hence, hybrid sensing is able to remarkably increase the \emph{density} of geographic information. As a consequence, it is an effective technique for increasing the available volume of geographic information, while keeping the amount of messages to crawl manageable.

Figure~\ref{fig:dist-place-norm} shows the \emph{place variety} per event for solicited replies (\emph{reply2geo}) and spontaneous \emph{relevant} tweets. Results show that hybrid sensing improved the \emph{variety} of geographic information up to $18\times$. Moreover, Figure~\ref{fig:new-place-norm} reports the \emph{coverage gain} obtained by applying the hybrid sensing approach with respect to the traditional opportunistic one. As can be seen, the geographic \emph{coverage} of the system increased up to $30\%$. Notably, these results may enable to fill the information gaps related to otherwise missing or underrepresented areas. For example, it can contribute to avoid sparse crisis maps~\cite{avvenuti2018crismap}.

Another interesting improvement concerns the \emph{granularity distribution} of the places mentioned in tweets. In fact, a geotag at a fine-grained level (e.g.: \emph{building}) is more valuable than one at a coarse-grained level (e.g.: \emph{country}) in a situational awareness perspective, in which information enrichment is intended to produce actionable knowledge~\cite{middleton2014real}. We computed the granularity of a place by leveraging our \emph{geo-semantic-parsing}  technique~\cite{avvenuti2018gsp}, that exploits the \texttt{rdf:type} predicates of the geographic resources. Figure~\ref{fig:place-types} outlines the overall distribution of the granularity of extracted places. Place types are ordered from finer (\emph{building}) to coarser (\emph{other}) grains. Clearly, solicited replies contained finer-grained geographic information with respect to spontaneous messages: we observed less instances of the most generic \emph{others} type, as opposed to $5.4\times$ more \emph{regions} and $2.5\times$ more \emph{buildings}. As a result, HERMES hybrid sensing was able to improve the granularity of the geographic information with respect to traditional, opportunistic approaches.

 \section{Conclusions and future work}
\label{sec:conc}
 
We described a real-world experiment with a novel system based on hybrid sensing and relying on a collection of state-of-the-art AI techniques. The aim was to demonstrate the feasibility of rapidly collecting and improving the quality of social data in the aftermath of mass emergencies. Among the peculiarities of our system is the possibility to be configured according to the information needs of EOCs. With a long-lived sensing campaign on Twitter, we demonstrated the practical usefulness of our solution, that allowed us to collect an averaged $+20\%$ additional data. \hl{Notably, the newcomer HERMES was able to engage more participants than the well-established USGS-DYFI on earthquakes occurring outside the U.S. territory and receiving little media coverage.} Furthermore, such data is richer in information than that typically used in OSN-based disaster management, as we showed that it is possible to increase the volume, variety, and the granularity of social crisis data in a timely fashion.

The compelling, yet preliminary results of our study also pave the way for future research and experimentation. Among the desirable outcomes of our work is the possibility to embed the hybrid sensing paradigm into existing disaster management systems. For example, the combination of machine and human intelligence exploited in systems like AIDR \cite{imran2014aidr} should enable a better matching between the information needs of EOCs and social data and, in general, allow us to exploit on-the-ground witnesses for obtaining the best possible information. In this work we have just faced this challenge, which indeed requires further and more in-depth studies.

 \section*{Acknowledgements}

This research is supported in part by the EU H2020 Program under the schemes \texttt{INFRAIA-1-2014-2015: Research Infrastructures} grant agreement \#654024 \textit{SoBigData: Social Mining \& Big Data Ecosystem}. 

\bibliography{references}

\end{document}